\RequirePackage{lineno}
\documentclass[aip,pop,groupedaddress,reprint]{revtex4-1}
\usepackage[dvipdfm]{graphicx}
\usepackage{dcolumn}
\usepackage{bm}

\begin{document}

\title{Dynamics and microinstabilities at perpendicular collisionless shock:
A comparison of large-scale two-dimensional full particle simulations 
with different ion to electron mass ratio}

\author{Takayuki Umeda}
\email[Email:]{umeda@stelab.nagoya-u.ac.jp}
\affiliation{Solar-Terrestrial Environment Laboratory, 
Nagoya University, Nagoya 464-8601, JAPAN}

\author{Yoshitaka Kidani}
\affiliation{Solar-Terrestrial Environment Laboratory, 
Nagoya University, Nagoya 464-8601, JAPAN}

\author{Shuichi Matsukiyo}
\email[Email:]{matsukiy@esst.kyushu-u.ac.jp}
\affiliation{Earth System Science and Technology, 
Kyushu University, Kasuga 816-8580, JAPAN}

\author{Ryo Yamazaki}
\email[Email:]{ryo@phys.aoyama.ac.jp}
\affiliation{Department of Physics and Mathematics, 
Aoyama Gakuin University, Sagamihara 252-5258, JAPAN}


\begin{abstract}
Large-scale two-dimensional (2D) full particle-in-cell simulations 
are carried out for studying the relationship between 
the dynamics of a perpendicular shock and 
microinstabilities generated at the shock foot. 
The structure and dynamics of collisionless shocks are generally 
determined by Alfven Mach number and plasma beta, while 
microinstabilities at the shock foot are controlled by 
the ratio of the upstream bulk velocity to 
the electron thermal velocity and the ratio of the plasma-to-cyclotron frequency. 
With a fixed Alfven Mach number and plasma beta, 
the ratio of the upstream bulk velocity to 
the electron thermal velocity is given as a function of 
the ion-to-electron mass ratio. 
The present 2D full PIC simulations 
with a relatively low Alfven Mach number ($M_A \sim 6$) show that 
the modified two-stream instability is dominant 
with higher ion-to-electron mass ratios. 
It is also confirmed that 
waves propagating downstream are more enhanced 
at the shock foot near the shock ramp
as the mass ratio becomes higher. 
The result suggests that these waves 
play a role in the modification of the dynamics of collisionless shocks 
through the interaction with shock front ripples. 
\end{abstract}

\pacs{
52.35.Tc; 
52.35.Qz; 
52.65.-y; 
52.65.Rr; 
}


\maketitle


\section{Introduction}

Collisionless shocks have been investigated by 
full Particle-In-Cell (PIC) simulations 
for more than four decades since early 1970's \cite{Biskamp_1972}. 
The full PIC method handles 
both electron-scale microphysics and ion-scale shock nonstationarity 
(i.e., spatiotemporal variation) simultaneously, 
since both electrons and ions are treated as individual charged particles. 
However, smaller simulation systems and 
reduced parameters (such as the ion-to-electron mass ratio $m_i/m_e$, 
the electron plasma-to-cyclotron frequency ratio $\omega_{pe}/\omega_{ce}$, 
and the speed of light relative to the electron thermal velocity ratio $c/v_{te}$) 
were used in past simulation studies to save the computational cost. 

Since early one-dimensional (1D) simulations 
\cite[e.g.,][]{Biskamp_1972,Quest_1985,Lembege_1987}, 
it has been well known that the shock front at 
supercritical (quasi-)perpendicular collisionless shocks 
becomes nonstationary. 
Incoming ions are decelerated and accumulated at the shock front. 
Then, a part of them are reflected upstream periodically. 
This coherent behavior of incoming ions results in 
the periodic collapse and redevelopment of the shock front, 
which is called the self-reformation. 
Later, the existence of the reformation 
in the two-dimensional (2D) system
was also confirmed by a full PIC simulation 
\cite{Lembege_1992}. 

It has been known that various types of microinstabilites are 
generated at the shock foot of 
perpendicular and quasi-perpendicular shocks 
during the broadening phase of the shock reformation 
when a part of incoming ions are reflected upstream. 
The ion reflection results in the deceleration of incoming electrons 
so that the conservation of the total current 
(the zero current condition in the shock normal direction) is satisfied. 
Consequently, there arises a relative drift velocity 
between the incoming electrons and the incoming/reflected ions, 
which is the free energy source of these microinstabilities. 
The ratio of the relative drift velocity 
to the electron thermal velocity is important for 
controlling the type of microinstabilities 
(or the effect of electron thermal damping to waves) 
\cite{Scholer_2003,Scholer_2004}. 
This ratio is proportional to 
the ratio of the upstream bulk velocity 
to the electron thermal velocity. 
Past self-consistent kinetic simulation studies revealed that
there exist various types of microinstabilities in the shock foot region of 
perpendicular and quasi-perpendicular shocks. 

Structures and dynamics of shock waves are generally 
determined by the following two dimensionless parameters, i.e., 
Alfven Mach number 
\begin{equation}
\label{eq:mach}
M_A = \frac{u_{1}}{c}\frac{\omega_{pi1}}{\omega_{ci1}}
    = \frac{u_{1}}{c}\frac{\omega_{pe1}}{\omega_{ce1}}\sqrt{\frac{m_i}{m_e}},
\end{equation}
and the plasma beta 
(the ratio of the thermal plasma pressure to the magnetic pressure)
\begin{equation}
\beta_e = \frac{2v_{te1}^2\omega_{pe1}^2}{c^2\omega_{ce1}^2}, \ \ \ 
\beta_i = \frac{2v_{ti1}^2\omega_{pi1}^2}{c^2\omega_{ci1}^2}. 
\end{equation}
where $c$, $u$, $\omega_{p}$, $\omega_{c}$ and $v_{t}$ 
represent the speed of light, bulk velocity, 
plasma frequency, cyclotron frequency and thermal velocity, 
respectively, with the subscripts ``$i$'' and ``$e$'' 
being ion and electron, respectively. 
Here the subscript ``1'' denotes the upstream. 
From these definitions of the Alfven Mach number and the electron beta, 
we obtain the ratio of the upstream bulk velocity 
to the electron thermal velocity, 
which determines the type of microinstabilities in the shock foot region, 
\begin{equation}
\frac{u_{x1}}{v_{te1}} = 
\frac{\sqrt{2}M_A}{\sqrt{\beta_e}}\sqrt{\frac{m_e}{m_i}}.
\end{equation}
This relation means that 
the ratio of the upstream bulk velocity to 
the electron thermal velocity becomes larger 
with larger Alfven Mach number, smaller electron beta, 
or smaller mass ratio. 
Note that the actual amount of free energy 
relative to the electron thermal energy 
($m_i u_{x1}^2/m_e v_{te1}^2$) is 
independent of the mass ratio. 

The relative bulk velocity between electrons and ions 
(which is proportional to $u_{x1}/v_{te1}$)
controls the type of microinstabilities, 
while the ratio of the plasma-to-cyclotron frequency ($\omega_{pe}/\omega_{ce}$)
determines the dispersion relation of plasma waves. 
When the relative bulk velocity between electrons 
and ions exceeds the electron thermal velocity, 
electrostatic waves are excited by current-driven instabilities 
such as the Buneman-type instability (BI) \cite{Buneman_1958} or 
the electron cyclotron drift instability (ECDI) \cite{Wong_1970,Forslund_1970}. 
At high-Mach-number perpendicular shocks, 
the relative velocity between incoming electrons and 
reflected ions commonly becomes much faster 
than the electron thermal velocity. 
Then, the BI becomes dominant, and 
electrostatic waves are excited 
at the upper hybrid resonance frequency  
\cite{Shimada_2000}. 
At lower-Mach-number ($M_A < 10$) perpendicular shocks, 
the relative velocity between incoming electrons and 
incoming/reflected ions becomes close to 
the electron thermal velocity. 
Then, the growth rate of the BI 
becomes small because of the damping by thermal electrons, 
and the ECDI becomes dominant, which excites 
electrostatic waves at 
multiple electron cyclotron harmonic frequencies 
\cite{Muschietti_2006}. 
When the relative velocity between incoming electrons and 
incoming/reflected ions becomes slower 
than the electron thermal velocity 
at lower-Mach-number perpendicular shocks, 
high-frequency electrostatic waves are not excited 
due to the damping by thermal electrons, and 
the modified two-stream instability (MTSI) 
\cite{Lashmore_1971,Krall_1971,Ott_1972,McBride_1972a,McBride_1972b} 
becomes dominant. 
Then, obliquely propagating electromagnetic whistler mode waves 
are excited at a frequency 
between the electron cyclotron frequency and 
the lower hybrid resonance frequency  
\cite{Wu_1983,Scholer_2003,Matsukiyo_2003,Scholer_2004,
Matsukiyo_2006,Matsukiyo_2010,Umeda_2012a}. 
Our previous study \cite{Umeda_2012b} has clearly shown that 
for a relatively low Mach number ($M_A=6$) perpendicular shock 
the MTSI becomes dominant with higher mass ratios ($m_i/m_e \ge 100$) 
while the ECDI becomes dominant with smaller mass ratio ($m_i/m_e=25$).

Meanwhile, 
it has been also known by the early hybrid PIC simulations 
\cite[e.g.,][]{Winske_1988} 
that large-amplitude fluctuations commonly exist 
along the shock overshoot in multidimensional systems, 
which is called as the ``ripples.'' 
A typical spatial scale of the shock-front ripples 
is $4-8$ ion inertial lengths 
in the shock-tangential direction. 
The ripples are thought to involve (L-mode) Alfven ion cyclotron waves 
excited by ion temperature anisotropy at the shock front \cite{Winske_1988}. 
So far, it has been difficult to perform 
2D full PIC simulations including the shock-front ripples, 
since current computer resources are not necessarily 
enough to cover a large simulation domain of 
several ion inertial lengths in the shock-tangential direction. 

Development of computer technologies in recent days 
allows us to perform higher-resolution hybrid 2D PIC  
and larger-scale 2D full PIC simulations with a longer simulation time. 
Recent 2D full PIC and hybrid PIC simulations 
of exactly perpendicular shocks 
\cite{Hellinger_2007,Lembege_2009} 
have suggested that the dynamics of collisionless shocks 
is modified by the ion-to-electron mass ratio. 
It has been demonstrated that the reformation is absent
with the ion-to-electron mass ratio $m_i/m_e=400$, 
while the reformation is suppressed with $m_i/m_e=42$ 
where the shock reformation is evident in an early phase 
but becomes apparently suppressed in a later phase. 
The difference in the shock dynamics with different mass ratios 
in the previous studies 
was discussed in terms of the excitation of 
``nonlinear whistler waves'' at the shock front. 


Our previous study \cite{Umeda_2012b} has confirmed that 
the mass ratio controls microinstabilities in the shock foot region. 
However, the generation of the ripples 
at the shock front was not included in our previous study, 
because the system length in the shock-tangential direction was too short. 
The purposes of the present study are 
to perform a large-scale 2D simulation including shock-front ripples 
for understanding the competition between the ripples and microinstabilities 
and to make a direct comparison between 
simulation results with different ion-to-electron mass ratio. 
It should be noted that the existence of the shock reformation 
itself \cite{Yuan_2009} is not discussed in the present study. 

The paper is organized as follows.
Section II describes the model and the parameters 
of full PIC simulations.
Section III shows the simulation results, 
and section IV gives conclusion and discussion of the present study.

\section{Simulation Setup}

We use a 2D electromagnetic full particle code 
in which the full set of Maxwell's equations and 
the relativistic equation of motion for individual electrons and ions 
are solved in a self-consistent manner. 
The continuity equation for charge is also 
solved to compute the exact current density 
given by the motion of charged particles \cite{Umeda_2003}. 
In the present simulation, 
the simulation domain is taken in the shock-rest frame 
\cite[e.g.,][]{Leroy_1981,Leroy_1982,Muschietti_2006,Umeda_2006}. 
A collisionless shock is excited by the ``relaxation'' 
between a supersonic plasma flow and 
a subsonic plasma flow moving in the same direction. 
The detailed initial setup is described in Refs. 
\cite{Umeda_2008,Umeda_2009}. 

In the present study, 
we assume a low-Mach number ($M_A\sim6$), 
moderate beta ($\beta_{i1}=\beta_{e1}=0.32$), 
weakly-magnetized 
($\omega_{pe1}/\omega_{ce1}=4$), and 
perpendicular ($\theta_{B_n} =90^{\circ}$) collisionless shock 
as shown in Table 1. 
Here, subscripts ``1'' and ``2'' denote 
``upstream'' and ``downstream'', respectively. 
We take the simulation domain in the $x$-$y$ plane 
and assume an in-plane shock magnetic field ($B_{y0}$). 
As a motional electric field, a uniform external electric field 
$E_{z0} = u_{x1}B_{y01} (= u_{x2}B_{y02})$ 
is applied in both upstream and downstream regions,  
so that both electrons and ions drift along the $x$ axis. 
At the left boundary of the simulation domain in the $x$ direction, 
we inject plasmas with the same quantities 
as those in the initial upstream region, 
while plasmas with the same quantities as those 
in the initial downstream region are also injected from the right boundary 
in the $x$ direction. 
We use absorbing boundaries 
to suppress non-physical reflection of electromagnetic waves at 
both ends of the simulation domain in the $x$ direction 
\cite{Umeda_2001}, 
while the periodic boundaries are imposed in the $y$ direction.

\begin{table*}[b]
\caption{
Simulation parameters used by different authors. 
}
\begin{tabular}{l||c|c|c|c|c|c|c}
Authors & Code & $M_A$ & $\theta_{B_n}$ & $\beta_{i1}$ & $\beta_{e1}$ & 
 $m_i/m_e$ & $\omega_{pe1}/\omega_{ce1}$ 
\\ \hline 
Run A (Present) & full PIC &
6.5 & $90^{\circ}$ & 0.32 & 0.32 & 25   & 4 \\
Run B (Present) & full PIC &
6.58 & $90^{\circ}$ & 0.32 & 0.32 & 100  & 4 \\
Run C (Present) & full PIC &
6.58 & $90^{\circ}$ & 0.32 & 0.32 & 256  & 4 \\
Run D (Present) & full PIC &
6.58 & $90^{\circ}$ & 0.32 & 0.32 & 625  & 4 \\
\hline
\textit{Hellinger et al.} \cite{Hellinger_2007} & Hybrid &
3.6 & $90^{\circ}$ & 0.2   & 0.5   & $-$ & $-$ \\
\textit{Lembege et al.} \cite{Lembege_2009} & full PIC &
4.93& $90^{\circ}$ & 0.15  & 0.24  & 400 & 2 \\
\ \ \ \ \ \ --  & full PIC &
4.93& $90^{\circ}$ & 0.15  & 0.24  & 42 & 2 \\
\textit{Umeda et al.} \cite{Umeda_2009,Umeda_2010,Umeda_2011} & full PIC &
5.5 & $90^{\circ}, 80^{\circ}$ & 0.125 & 0.125 & 25  & 10 \\
\end{tabular}
\end{table*}

We performed four simulation runs (A, B, C and D) 
with different ion-to-electron mass ratio $m_i/m_e=$ 25, 100, 256 and 625, 
respectively. 
The grid spacing and the time step of the present simulation runs are 
set to be $\Delta x = \Delta y \equiv \Delta = \lambda_{De1}$ 
and $c\Delta t/\Delta = 0.5$. 
Here $\lambda_{De}$ is the electron Debye length. 
The total size of the simulation domain is 
$32 l_{i1} \times 6 l_{i1}$, 
where $l_{i1} = c/\omega_{pi1}$ is the ion inertial length 
($=50\lambda_{De1}$, $=100\lambda_{De1}$, $=160\lambda_{De1}$ 
and $=250\lambda_{De1}$ 
in Runs A, B, C and D, respectively). 
The bulk flow velocity of the upstream plasma is 
$u_{x1}/v_{te1} = $ 3.0, 1.5, 0.9375 and 0.6 in Runs A, B, C and D, 
respectively ($u_{x1}/V_{A1} = 6$), which controls the free energy source 
for microinstabilities in the shock foot region. 
Note that the previous small-scale 2D full PIC simulations 
(without ripples) \cite{Umeda_2012b} 
have confirmed that the ECDI is dominant with the parameters for Run A 
while the MTSI is dominant with the parameters for Runs B--D. 


The initial state consists of two uniform regions 
separated by a discontinuity. 
In the upstream region that is taken in the left-hand side 
of the simulation domain, 
electrons and ions are distributed uniformly in space and 
are given random velocities $(v_x,v_y,v_z)$ to approximate 
shifted Maxwellian momentum distributions 
with the drift velocity $\boldmath{u}_1$, 
number density $n_{1} \equiv \epsilon_0 m_e \omega_{pe1}^2 / e^2$, 
isotropic temperatures $T_{e1} \equiv m_e v_{te1}^2$ and 
$T_{i1} \equiv m_i v_{ti1}^2$, 
where $m$, $e$, $\omega_{p}$ and $v_{t}$ are 
the mass, charge, plasma frequency and 
thermal velocity, respectively. 
The upstream magnetic field $\boldmath{B}_{01}$ 
with a magnitude of $m_e \omega_{ce1}/e$ 
is also assumed to be uniform, where $\omega_{c}$ 
is the cyclotron frequency. 
The downstream region taken in the right-hand side 
of the simulation domain is prepared similarly with 
the drift velocity $\boldmath{u}_2$, density $n_{2}$, 
isotropic temperatures $T_{e2}$ and $T_{i2}$, 
and magnetic field $\boldmath{B}_{02}$. 

In the relaxation method, 
the initial condition is given by solving 
the shock jump conditions (Rankine-Hugoniot conditions) for 
a magnetized two-fluid isotropic plasma 
consisting of electrons and ions \cite{Hudson_1970}. 
In order to determine a unique initial downstream state, 
we need given upstream quantities 
$u_{x1}$, $\omega_{pe1}$, $\omega_{ce1}$, $v_{te1}$, 
$v_{ti1}$, and $T_{i2}/T_{e2}$. 
Note that the initial downstream ion-to-electron temperature ratio 
$T_{i2}/T_{e2} = 8.0$ is assumed 
to solve the jump condition for a unique initial quantities 
of the downstream plasma. 
It should be noted that a shock wave is excited by the relaxation of 
the two plasmas with different quantities
in the present shock-rest-frame model. 
Since the initial state is given by the shock jump conditions 
for a ``two-fluid'' plasma consisting of electrons and ions \cite{Hudson_1970}, 
however, 
the kinetic effect is excluded in the initial state 
and it is very difficult to construct an exact shock rest frame. 
Then, the shock front of the excited shock moves upstream 
at a slow velocity ($v_{sh} \sim 0.5-0.6V_{A1}$), as we will show later. 
Thus, the Mach number should be corrected as shown in Table 1. 

We used 25 pairs of electrons and ions per cell in the upstream region 
and 64 pairs of electrons and ions per cell in the downstream region, 
respectively, at the initial state.

\section{Results}

\begin{figure*}[t]
\center
\includegraphics[width=1.0\textwidth,bb=0 0 1900 900]{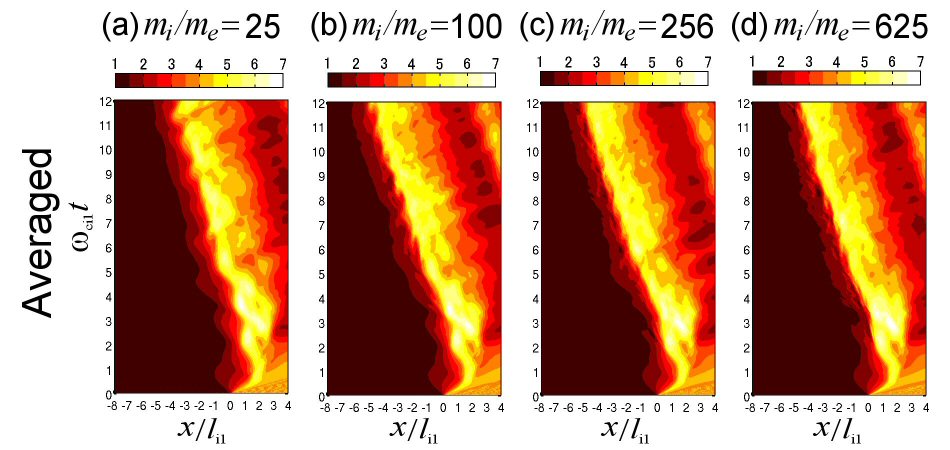}
\caption{
Tangential magnetic field $B_y$ 
as a function of position $x$ and time $t$ 
for Runs A--D. 
The position and time are normalized by 
$l_{i1}$ and $1/\omega_{ci1}$, 
respectively. 
The magnitude is 
normalized by the initial upstream magnetic field $B_{y01}$. 
The averaged magnetic field over the $y$ direction is shown.  
}
\end{figure*}

Figure 1 shows the tangential component of 
the shock magnetic field ($B_y$) for Runs A--D 
as a function of position $x$ and time $t$. 
The position and time are renormalized by 
the ion inertial length $l_{i1} = c/\omega_{ci1}$ and 
the ion cyclotron angular period $1/\omega_{ci1}$, respectively. 
The magnitude is normalized by the 
initial upstream magnetic field $B_{01}$. 
In Fig.1, the magnetic field $B_y$ is 
``averaged'' over the $y$ direction, 
which means that fluctuations in the shock-tangential direction 
are smoothed out. 

In all the runs, 
the shock overshoot ($|B_y|\sim 6.5$) appears at 
$\omega_{ci} t \sim 2.2$ and 3.6. 
After that, however, these runs show different development. 
In Run A, 
a strong magnetic field ($|B_y|\sim 6$) appears 
at $\omega_{ci} t \sim 5.5$ and 6.8, 
and then the oscillation becomes smaller and smaller. 
This is similar to the previous study\cite{Lembege_2009}, which reported  
the (nonphysical) suppression of the reformation 
in the result with $m_i/m_e=42$. 
It is noted that the previous study \cite{Umeda_2010}  with $m_i/m_e=25$
demonstrated that the quasi-periodic oscillation of the shock overshoot gradually
disappears when the magnetic field is averaged over the shock tangential direction. 
However, this fact does not necessarily mean the
non-existence of the shock reformation.

In Run B, 
a strong magnetic field ($|B_y|\sim 6$) appears 
at $\omega_{ci} t \sim 5.0$ and 7.5, 
and then the oscillation becomes smaller and smaller. 
There also appears a small-scale (smaller than the ion scale) structure 
at the shock front (shock foot). 
By contrast, 
a strong magnetic field ($|B_y|\sim 6$) appears 
at $\omega_{ci} t \sim 6.0$ in Run C and  
$\omega_{ci} t \sim 7.0$ in Run D, 
but the quasi-periodic oscillation of the shock overshoot  
is not evident. 
This is similar to the previous study\cite{Lembege_2009}, which reported  
the absence of the reformation 
in the result with $m_i/m_e=400$. 
There also appear smaller-scale structures 
at the shock front (shock foot) in Runs C and D.

\begin{figure*}[t]
\center
\includegraphics[width=1.0\textwidth,bb=0 0 3400 3550]{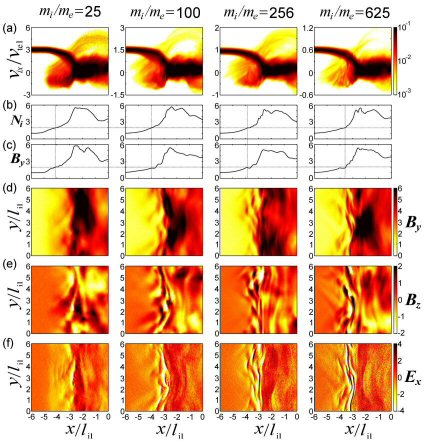}
\caption{
Typical structure of the perpendicular shock 
transition region for Runs A--D at a quasi-steady phase ($\omega_{ci}t=11$). 
(a) The $x-v_x$ phase space density of ions, 
(b) the ion density $N_i$, and (c) the shock magnetic field $B_y$ 
averaged over $y$. 
The spatial profile of (d) $B_y$, (e) $B_z$ and (f) $E_x$. 
The magnetic field is 
normalized by the upstream magnetic field $B_{y01}$, and 
the electric field is 
normalized by the motional electric field $E_{z0}$. 
}
\end{figure*}

\begin{figure*}[t]
\center
\includegraphics[width=1.0\textwidth,bb=0 0 3400 3550]{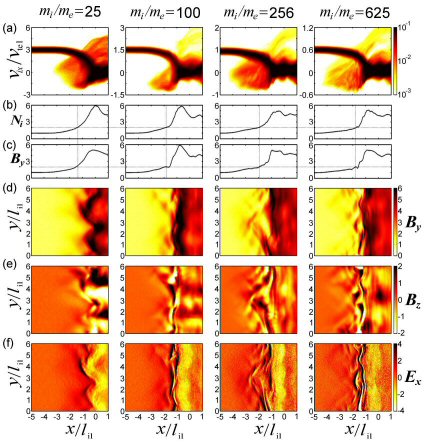}
\caption{
Structure of the perpendicular shock 
transition region for Runs A--D at a transient phase ($\omega_{ci}t=5.0$) 
with the same format as Fig.2. 
}
\end{figure*}

\begin{figure*}[t]
\center
\includegraphics[width=1.0\textwidth,bb=0 0 3400 3550]{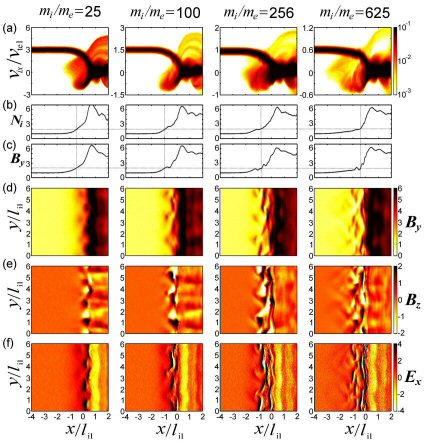}
\caption{
Structure of the perpendicular shock 
transition region for Runs A--D at an early phase ($\omega_{ci}t = 3.6$) 
with the same format as Fig.2. 
}
\end{figure*}

Figure 2 shows a typical structure of the perpendicular shock 
transition region for Runs A--D at a quasi-steady state ($\omega_{ci}t=11$). 
The panel (a) shows the $y$-averaged $x-v_x$ phase space density of ions. 
The panels (b) and (c) show the $y$-averaged 
ion density $N_i$ and shock magnetic field $B_y$, respectively. 
The panels (d),  (e) and (f) show the corresponding spatial profile 
of $B_y$, $B_z$ and $E_x$, respectively. 
In all the runs, a large-amplitude fluctuation with a wavelength of 
six ion inertial lengths exists in the magnetic field $B_y$ and $B_z$ components 
at the shock overshoot, which is identified as the ripples. 

In Run A, we can see the characteristic signature of 
the shock transition region, i.e., foot, ramp, and overshoot. 
The spatial size of the foot region in Run A 
corresponds to the gyro radius of reflected ions, 
which is $\sim l_{i1}$ in this case. 
The typical ion density and shock magnetic field 
in the foot region is $N_i \sim 2.5n_{1}$ and $B_y \sim 2.5B_{y01}$, 
respectively. 
The spatial profiles of $B_y$ and $B_z$ respectively in panels (d) and (e) 
show that there are large-scale (ion-scale) 
waves at the shock front, which correspond to the rippled structure. 
The spatial profile of $E_x$ in panel (f) shows 
the excitation of electron-scale electrostatic waves (at $y/l_{i1} \sim 3.5$). 
It is expected that these waves are due to the ECDI generated by 
a localized ion beam \cite{Umeda_2009,Umeda_2011}. 

The spatial profiles of $B_y$, $B_z$ and $E_x$ in Runs B--D show 
the excitation of electron-scale electromagnetic waves, 
which implies oblique whistler waves due to the MTSI 
\cite{Matsukiyo_2003,Matsukiyo_2006,Umeda_2012a}. 
The previous linear analysis \cite{Umeda_2012b} 
has shown that the wavelength of the whistler waves becomes shorter 
as the mass ratio becomes higher, which is consistent with the present study. 
On the other hand, the generation of the MTSI is not so clear in Run A. 

As the mass ratio becomes higher, 
the ion density and shock magnetic field in the foot region 
becomes smaller. 
The spatial size of the foot region also looks smaller 
as the mass ratio becomes higher. 
As seen in the panel (a) of Runs B--D, however, 
reflected ions are strongly scattered in the shock normal direction 
by the waves and spread over a wide range in the $v_x$ space, 
and a small part of reflected ions reach $x/l_{i1} = -6$.

Figure 1 indicates that the dynamics of the shock front looks similar in all the runs 
until $\omega_{ci}t \sim 4$ but becomes different after $\omega_{ci}t \sim 5$. 
Figure 3 shows a structure of the shock transition region for Runs A--D 
at a transient phase ($\omega_{ci}t=5.0$). 
There is a difference in the evolution of the shock front 
among these runs. 
In Run A, the shock front is kinking at Mode 2 in $y$. 
In Run B, there appears a fluctuation at Mode 3 in $y$, 
and in Runs C and D, 
the shock overshoot is modified at Mode 2 in $y$. 
At the shock foot, 
there also appear small-scale wave activities near the shock ramp in Runs B--D. 
These small-scale waves are more enhanced as the mass ratio becomes higher. 

Figure 4 shows a structure of the shock transition region for Runs A--D 
at an early phase ($\omega_{ci}t=3.6$), 
when the periodic self-reformation is seen in Run A. 
In all the runs, there is a quasi-1D structure 
of the negative electric field $E_x$ component (i.e., shock potential) 
at the shock overshoot, 
implying that shock-front ripples are not generated at this time. 
There also appears a fluctuation in the magnetic field $B_z$ component 
around the shock overshoot 
at Mode 5, 4, 3, and 3 in $y$ in Run A, B, C, and D, respectively, 
which corresponds to L-mode cyclotron waves \cite{Winske_1988} 
and may develop into shock-front ripples. 
At the shock foot, 
there appears a fluctuation at Mode 3 in $y$ ($k_yl_{i1} \sim 2.1$) 
and $x/l_{i1} \sim -0.5$ in Runs B and C. 
However, the fluctuation at Mode 3 in $y$ is not evident in Run D 
but small-scale fluctuations with various mode numbers are seen. 
The small-scale wave activities are also seen in Run C near the shock ramp, 
suggesting that the small-scale fluctuations are more enhanced 
as the mass ratio becomes higher. 
In Run A, 
there appears a fluctuation at Mode 3 in $y$ ($k_yl_{i1} \sim 2.1$) 
around the shock ramp ($x/l_{i1}\sim 0.2$).  

Note that the excitation of waves due to the relative drift 
between ions and electrons at the shock foot region is 
discussed by many authors. 
It is suggested 
that there are two possible excitation types of 
instability at the shock foot 
due to incoming ions and reflected ions, 
which are referred as ``instability-1'' ($k_x > 0$) and ``instability-2'' ($k_x < 0$), respectively 
\cite{Matsukiyo_2006}. 
The reflected ion component in the phase space plots in Figs3(a) and 4(a) 
shows phase-space vortices in Runs B--D, 
which indicate the excitation of waves due to microinstabilites 
(possibly by the MTSI-2). 
Simultaneously, the incoming ion component also shows modulation near the shock ramp 
in Runs B--D.

Our previous studies \cite{Umeda_2012a,Umeda_2012b} presented 
a series of small-scale 2D full PIC simulation with the same physical parameters 
but for different mass ratio. 
Also, a direct comparison was made between the simulation result and 
a linear analysis of waves excited at the shock foot  
due to incoming and reflected ions. 
An excellent agreement 
between the linear analysis and the simulation result was shown. 
However, since the simulation domain was 
shorter than the ion inertial length in the shock-tangential direction, 
shock-front ripples were not included. 
In the present large-scale simulations, by contrast, 
since shock front ripples generate an inhomogeneous reflected ion beam 
\cite{Umeda_2009,Umeda_2011},
the standard linear analysis 
(in which the spatial distribution of plasma is assumed to be uniform) 
is not directly applicable. 
However, 
we use the result of previous linear analysis \cite{Umeda_2012a,Umeda_2012b} 
as a reference of microinstabilites at the shock foot.

Figures 5, 6 and 7 show 
the spectral intensity of the electric field $E_x$ component 
in the shock foot region at different time intervals 
($-2 \le x/l_i < 0$ and $0 \le y/l_i < 6$ for a time interval of $\omega_{ci} t = 3.5-4.5$ for Fig.5, 
$-3 \le x/l_i < -1$ and $0 \le y/l_i < 6$ for a time interval of $\omega_{ci} t = 5-6$ for Fig.6, and  
$-5 \le x/l_i < -3$ and $0 \le y/l_i < 6$ for a time interval of $\omega_{ci} t = 10-11$ for Fig,7, 
respectively). 
The $k_x-k_y$ spectrum in the top panels and 
$\omega-k_x$ spectrum in the bottom panels are 
obtained by integrating the $\omega-k_x-k_y$ spectrum 
over $\omega$ and $k_y$, respectively. 
The wave intensity is normalized by the motional electric field $E_{z0}$. 
The gradient of the dashed line indicates the speed of the shock wave 
($-0.5V_{A1} = -0.1l_{e1}\omega_{ce1}$ for Run A, 
$-0.58V_{A1} = -0.058l_{e1}\omega_{ce1}$ for Run B, 
$-0.58V_{A1} = -0.0363l_{e1}\omega_{ce1}$ for Run C, and 
$-0.58V_{A1} = -0.0232l_{e1}\omega_{ce1}$ for Run D, 
respectively). 
Note that the $\omega-k_x$ spectra in Fig.5 show more blurred feature
than those of Fig.3 in Ref.\cite{Umeda_2012b} 
due to the effect of integration in the larger region of $y$.

At an early phase ($\omega_{ci} t = 3.5-4.5$) in Fig.5, 
waves are generated in the $-x$ direction (toward upstream). 
These waves are enhanced below the electron cyclotron frequency 
and above the lower hybrid resonance frequency 
($\omega_{LHR}/\omega_{ce1} = $ 0.2, 0.1, 0.06 and 0.04 for Runs A, B, C and D, 
respectively). 
The phase velocities of these waves are faster than the speed of the shock wave 
($v_p \sim -0.44l_{e1}\omega_{ce1} = -1.1V_{te1}$ for Run A, 
 $v_p \sim -0.123l_{e1}\omega_{ce1} = -0.31V_{te1}$ for Run B, 
 $v_p \sim -0.1l_{e1}\omega_{ce1} = -0.25V_{te1}$ for Run C, and 
 $v_p \sim -0.085l_{e1}\omega_{ce1} = -0.21V_{te1}$ for Run D, respectively). 
These phase velocities are close to (about 50--70\% of) 
the drift velocity of reflected ions identified from Fig.4a.  
The previous linear analysis \cite{Umeda_2012b} 
suggested that the MTSI-2 due to reflected ions 
is unstable for a various mass ratio at almost the same electron-scale wavenumber 
in the shock-normal direction $k_x l_{e1} \sim -1.2$ 
and the same ion-scale wavenumber in the shock-tangential direction $k_y l_{i1} \sim 3.5$ 
(see Fig.5 in Ref.\cite{Umeda_2012b}). 
The $k_x-k_y$ spectra in Fig.5 show that 
these waves are enhanced at $k_x l_{e1} \sim -1$ and at $k_y l_{i1} \sim 3.14$ 
($k_yl_{e1} \sim $ 0.31, 0.2 and 0.13 for Runs B, C and D, respectively, 
which correspond to Mode 3 in $y$). 
It is confirmed that the oblique whistler waves are excited 
through the MTSI-2 due to reflected ions.

It is noted that the oblique whistler waves due to the MTSI-2 (at the shock foot) 
propagate at the phase velocity close to the drift velocity of reflected ions, 
while the L-mode cyclotron waves (at the shock ramp/overshoot) 
propagate at the phase velocity close to the propagation velocity of the shock wave. 
However, it is difficult to separate these waves in Run A 
because of the resolution in frequency space. 
In Run A, 
another high-frequency wave modes are also enhanced weakly 
above the electron cyclotron frequency. 
The phase velocity of this wave mode is much faster than the speed of the shock wave 
($v_p \sim -0.5l_{e1}\omega_{ce1} = -1.25V_{te1}$), 
which is close to the drift velocity of reflected ions.  
The result suggests the generation of ECDI-2 due to reflected ions. 
Note that ECDI-2 was confirmed in the previous small-scale simulation study 
with $m_i/m_e=25$ \cite{Umeda_2012b}, 
where electron cyclotron harmonic waves are enhanced at $\omega/\omega_{ce} \sim 3.5$ 
as predicted by the linear analysis (see Fig.3 in Ref.\cite{Umeda_2012b}).  
In the present large-scale simulation, on the other hand, 
the wave enhancement is not so strong and 
the spectrum spreads over a wide frequency range ($\omega \sim 1-3$). 
Here, the simulation result suggests the generation of the ECDI-2 is weakened 
by the modification of the linear dispersion relation due to  
the spatial inhomogeneity of the ion density.

It is also noted that 
in our previous small-scale 2D simulation 
(where shock-front ripples are not included), 
waves are strongly excited through the MTSI-2 (and the ECDI-2) 
but less through the MTSI-1 \cite{Umeda_2012b}. 
In the present large-scale simulation (with shock-front ripples), by contrast, 
waves with $k_x>0$ are also weakly excited at the early phase ($\omega_{ci} t = 3.5-4.5$), 
which is possibly by the MTSI-1 due to incoming ions 
as indicated from the modification of the phase-space distribution of reflected ions 
in Fig.4(a).

\begin{figure*}[t]
\center
\includegraphics[width=1.0\textwidth,bb=0 0 2420 560]{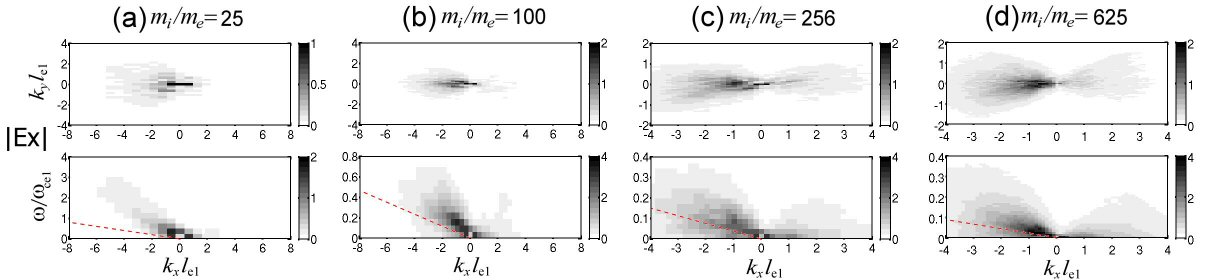}
\caption{
Spectral wave energy of the electric field $E_x$ component 
at the shock front ($-2 \le x/l_i < 0$ and $0 \le y/l_i < 6$)  
for a time interval of $\omega_{ci} t = 3.5-4.5$. 
The $k_x-k_y$ spectrum in the top panels and 
$\omega-k_x$ spectrum in the bottom panels are 
obtained by integrating the $\omega-k_x-k_y$ spectrum 
over $\omega$ and $k_y$, respectively. 
The wave intensity is normalized by the motional electric field $E_{z0}$. 
The dashed line indicates the speed of the shock wave. 
}
\end{figure*}

\begin{figure*}[t]
\center
\includegraphics[width=1.0\textwidth,bb=0 0 2420 560]{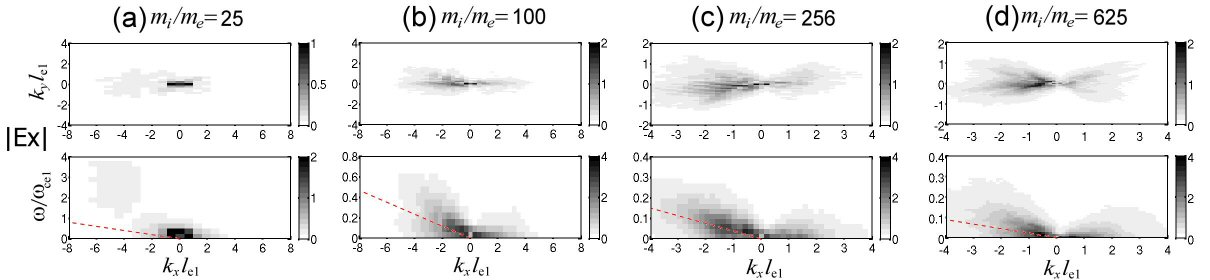}
\caption{
Spectral wave energy of the electric field $E_x$ component 
at the shock front ($-3 \le x/l_i < -1$ and $0 \le y/l_i < 6$)  
for a time interval of $\omega_{ci} t = 5-6$
with the same format as Fig.5. 
}
\end{figure*}

\begin{figure*}[t]
\center
\includegraphics[width=1.0\textwidth,bb=0 0 2420 560]{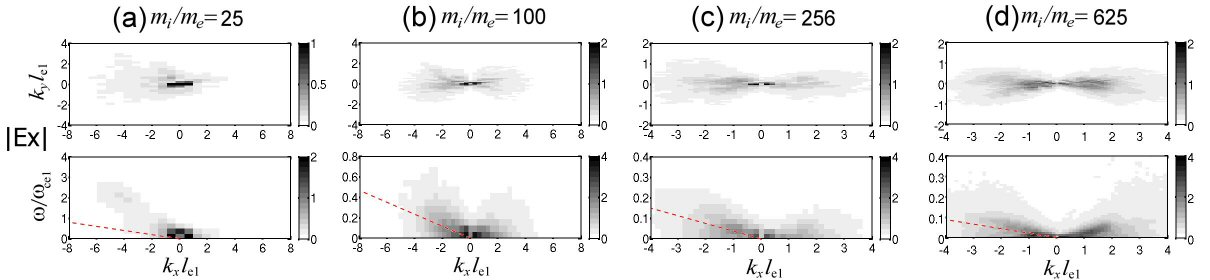}
\caption{
Spectral wave energy of the electric field $E_x$ component 
at the shock front ($-5 \le x/l_i < -3$ and $0 \le y/l_i < 6$)  
for a time interval of $\omega_{ci} t = 10-11$
with the same format as Fig.5. 
}
\end{figure*}

At a transient phase ($\omega_{ci} t = 5-6$) in Fig.6, 
waves propagating downstream are enhanced more than in Fig.5. 
For Runs B, C and D ($m_i/m_e = $ 100, 256 and 625), 
phase velocities of low-frequency waves propagating upstream 
become close to the propagation speed of the shock wave 
rather than the drift velocity of reflected ions. 
For Run A ($m_i/m_e = 25$), 
phase velocities of high-frequency waves propagating upstream are still close to the 
drift velocity of reflected ions. 
However, the generation of the ECDI-2 becomes less evident, 
which is because of the ``steepening phase'' of the shock reformation. 

The tendency of these wave characteristics becomes more evident 
at the quasi-steady phase ($\omega_{ci} t = 10-11$) in Fig.7. 
As the mass ratio becomes higher, 
waves propagating downstream (generated by the MTSI-1) are more enhanced. 
As the mass ratio becomes higher, 
waves propagating upstream at the drift velocity of reflected ions 
(generated by the MTSI-2) become less evident but 
waves propagating upstream at the speed of the shock wave 
becomes more evident. 
For Run A ($m_i/m_e = 25$), however, 
waves propagating upstream at the drift velocity of reflected ions 
(generated by the ECDI-2) are well identified.  
The agreement between the present large-scale simulation 
and the previous linear analysis becomes worse at a later phase 
because the coherent behavior of reflected ions is suppressed 
by the spatial inhomogeneity in higher mass-ratio runs. 
As seen in phase-space plots in Figs. 3(a) and 4(a) 
for Run C ($m_i/m_e = 256$) and Run D ($m_i/m_e = 625$), 
reflected ions are strongly scattered in the velocity space 
at the shock foot near the shock ramp, 
and the ``beam'' of reflected ions no longer exists.

\section{Conclusion and Discussion}

Our previous linear analysis and 
2D full PIC simulations with a small simulation domain \cite{Umeda_2012b} 
confirmed that the ion-to-electron mass ratio affects microinstabilities 
in the foot region of perpendicular collisionless shocks. 
However, the shock-front ripples were not included in the previous study. 
In the present study, we have performed 
2D full PIC simulations with a large simulation domain 
to include both shock front ripples and microinstabilities  
simultaneously. 

The previous linear analysis\cite{Umeda_2009,Umeda_2011} 
showed that the ECDI tends to be dominant 
with a low mass ratio, a low beta, or a low Mach number, 
and that the MTSI tends to be dominant 
with a high mass ratio, a high beta, or a low Mach number. 
The present large-scale 2D full PIC simulation shows 
a reasonable agreement with the linear analysis 
at an early phase of the simulation. 
However, the ECDI becomes weaker even for smaller mass ratio 
possibly due to the spatial inhomogeneity at the shock foot. 

By contrast to the reasonable agreement 
between the present large-scale simulation 
and the previous linear analysis at the early phase, 
the agreement becomes worse 
as both nonlinearity and spatial inhomogeneity at the shock front develop. 
The generation of the MTSI-2 becomes less evident for high-mass-ratio runs 
since reflected ions are scattered at the shock foot near the shock ramp 
by strong wave-particle interactions. 
Then, the phase velocity of waves propagating upstream switches to 
the propagation speed of the shock wave from the drift velocity of reflected ions.  
In addition to these results, 
it is also confirmed that waves propagating downstream are enhanced stronger 
at the shock ramp as time elapses and as the mass ratio becomes higher, 
possibly by the MTSI-1 due to incoming ions. 
The present result suggests that the difference of the shock dynamics 
between high and low mass ratios is caused by the waves propagating downstream.


Present study showed how the global shock structure 
and the wave excitation around shock front change with different mass ratio. 
However, the simulation data is not enough to give complete answers for 
the property and origin of these waves, unfortunately. 
Nevertheless, we discuss unsolved issues  
of the relationship between the shock dynamics and 
the wave excitation which we believe most likely.

The previous studies \cite{Hellinger_2007,Lembege_2009}
discussed the difference in the shock dynamics among different mass ratio 
in terms of the ``nonlinear whistler waves,'' which 
are excited at the shock front with $m_i/m_e =400$ 
but not with $m_i/m_e =42$. 
Is the present study with higher mass ratios 
consistent with the previous results \cite{Hellinger_2007,Lembege_2009}?
%
Hellinger \textit{et al.} \cite{Hellinger_2007} characterized the ``nonlinear whistler waves'' 
as oblique whistler waves at the shock foot with \emph{zero phase velocity in the shock rest frame}. 
Figure 2a in Ref.\cite{Hellinger_2007} shows that the waves with  
$k_{x}l_{i1} \sim 8$ and $k_{y}l_{i1} \sim 2.2$ are enhanced 
at both shock foot and shock ramp. 
Figure 2c in Ref.\cite{Lembege_2009} shows that 
waves with $k_{x}l_{i1} \sim 6$ and $k_{y}l_{i1} \sim 3$ 
are enhanced at the shock foot and 
waves with $k_{x}l_{i1} \sim 6$ and $k_{y}l_{i1} \sim 1.75$ 
are enhanced at the shock ramp. 
It is expected that the former one (at the shock foot) corresponds to 
the ``nonlinear whistler waves'' 
and that the latter one (at the shock ramp) corresponds to the shock-front ripples. 
In the present simulations, it is clearly shown by the $\omega-k_x$ spectra 
that there exist waves propagating at the propagating velocity of the shock wave. 
The $k_x-k_y$ spectrum at the quasi-steady phase 
(in Fig.7) shows that the waves are most enhanced 
at $k_{x}l_{i1} \sim 6$ and $k_{y}l_{i1} \sim 3.14$ for $m_i/m_e=$ 625 and 256 
($k_{x}l_{e1} \sim $ 0.25 and 0.4 for $m_i/m_e=$ 625 and 256, respectively). 
These characteristics are very similar to the ``nonlinear whistler waves'' 
seen in the previous studies \cite{Hellinger_2007,Lembege_2009}. 

However, the generation mechanism of the nonlinear whistler waves is still unclear. 
Although Hellinger \textit{et al.} \cite{Hellinger_2007} proposed a fluid-like 
nonlinear three-wave coupling, it is difficult to identify this process in the present study. 
As Hellinger \textit{et al.} \cite{Hellinger_2007} denied the direct generation 
by the beam of reflected ions (i.e., MTSI-2), 
reflected ions are strongly scattered in the velocity space 
at the shock foot near the shock ramp, possibly by the nonlinear whistler waves, 
and the beam of reflected ions no longer exists 
for higher-mass-ratio runs  in the present study. 
Here, we suggest the importance of the waves propagating downstream 
at the shock foot. 
As seen in the phase-space plots for higher-mass-ratio runs (Fig.2(a)), 
incoming ions are modulated and reflected ions are strongly scattered in the velocity space 
at the shock foot near the shock ramp. 
An alternative generation mechanism suggested in the present study 
is a kinetic wave-wave interaction in the phase space: 
the merging between phase-space vortices of large-amplitude (nonlinear) waves 
propagating upstream and downstream (generated by the MTSI-2 and MTSI-1, respectively), 
which may generate a large-amplitude phase-space vortices 
which is almost at rest in the shock frame. 
In order to confirm this process, however, a further investigation is required. 


Also, the issue of the absence/existence of the shock reformation 
discussed in the previous studies \cite{Hellinger_2007,Lembege_2009,Yuan_2009} 
is not addressed in the present study. 
This is because averaging the magnetic field over the shock tangential direction 
does not necessarily show the existence/absence of the shock reformation. 
Meanwhile, it is very difficult to define the shock reformation in multidimensional systems 
only from a local magnetic field \cite{Umeda_2010}. 
Here, we again suggest the importance of the waves propagating downstream 
at the shock foot. 
The waves generated by microinstabilities (MTSIs) have 
an electron-scale wavelength in the shock normal direction but 
an ion-scale wavelength in the shock tangential direction. 
Hence, it is possible that the waves propagating downstream generated by the MTSI-1 
interact with the L-mode Alfven cyclotron waves at the shock overshoot, 
which may result in the modification in the development of shock front ripples 
at the transition phase. 
Although this issue needs further parameter surveys 
for various Alfven Mach numbers and various plasma beta  
which should be left as a future study, 
the present result has indicated that 
microinstabilities play a role in the modification of 
collisionless shock dynamics among different ion-to-electron mass ratio.

\begin{acknowledgments}
This work was supported by MEXT/JSPS under 
Grant-in-Aid for Scientific Research on Innovative Areas No.21200050, 
Grant-in-Aid for Young Scientists (B) 
No.22740323 (S. M.) and No.21740184 (R. Y.). 
The computer simulations were performed on 
the DELL PowerEdge R815 supercomputer system at 
the Solar-Terrestrial Environment Laboratory (STEL) 
and the Fujitsu FX1 and HX600 supercomputer systems 
at the Information Technology Center (ITC), Nagoya University, 
as a STEL computational research program, 
a Nagoya University HPC research program, a JHPCN program 
at Joint Usage/Research Center for Interdisciplinary 
Large-scale Information Infrastructures, and 
a HPCI systems research project.  
\end{acknowledgments}

\end{document}